# STUDIES AND APPLICATION OF BENT CRYSTALS FOR BEAM STEERING AT 70 GEV IHEP ACCELERATOR

A.G. AFONIN, V.T. BARANOV, G.I. BRITVICH, V.N. CHEPEGIN,
YU.A. CHESNOKOV, V.I. KOTOV, V.A. MAISHEEV,
V.I.TEREKHOV, I.A. YAZYNIN,

*SRC IHEP,*
*Protvino, 142281,Russia*

This report overviews studies accomplished in the U70 proton synchrotron of IHEP-Protvino during the recent two decades. Major attention is paid to a routine application of bent crystals for beam extraction from the machine. It has been confirmed experimentally that efficiency of beam extraction with a crystal deflector of around 85% is well feasible for a proton beam with intensity up to $10^{12}$ protons per cycle. Another trend is to use bent crystals for halo collimation in a high energy collider. New promising options emerge for, say, LHC and ILC based on the "volume reflection" effect, which has been discovered recently in machine study runs at U70 of IHEP (50 GeV) and SPS of CERN (400 GeV).

## 1. Introduction

Ideas of use the particle channeling in bent crystals for steer the beams have been checked up and advanced in many experiments (see [1-3] and references herein). This method has found the widest practical application on U-70 accelerator of SRC IHEP, where crystals are used in regular runs for beam extraction and forming [4, 5].

## 2. Beam splitting

Usually a beam is split by an electrostatic or a magnetic splitter. This is a technically fairly complex approach requiring considerable space, since the angles of deflection of a beam by a conventional splitter are very limited. The use of crystals provides a simple means for beam splitting, which is unattainable by conventional techniques. The first crystal beam-splitting station [6] began to operate since 1988.

Recently in IHEP new splitting station has been created based on new-type bent crystal [7]. Its feature is the rational design of crystal bending device, allowed to minimize losses of particles at splitting. Due to commissioning of the new station 2 experimental setups could work simultaneously in different beam lines (see figure 1).

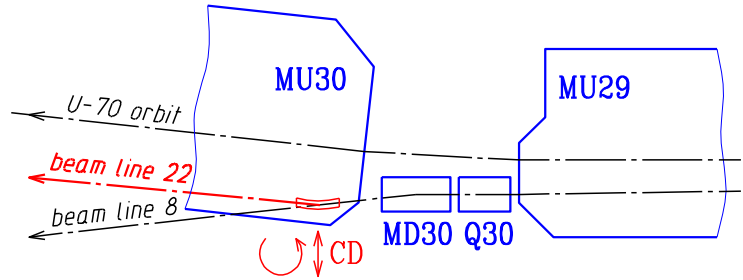

Figure 1. Layout of crystal splitting station.

Preparation of the bent crystal with use of two original methods has allowed lowering losses of particles at splitting up to 0.01 %, that on the order is better, than in the first experiments [6].

The first way consists in giving curvature with the help of making on one of surfaces of a crystal of microtrenches (about 20 micron depth), causing superficial zones with defects which create the total pressure, capable to bend a crystal (see [8]). Figure 2a explains the principle, and figure 2b shows realization of this method. Other way of a bend of a crystal plate is connected to use of anisotropic properties of a crystal lattice [9]. At a longitudinal bend of a plate (the figure 2c), arises an appreciable orthogonal bend which is realized in the device (a figure 2d). In the new device we have combined these both methods: the crystal with trenches on a surface was in addition bent in the device of type (figure 2d). Such approach provides high rate of a bend of a crystal, so the required size of a crystal along the beam was only 10 mm for providing 10 mrad bending angle. The new station of beam splitting worked about one month and has provided a beam with average intensity of $10^6$ particles in a second on a target of SVD experimental setup.

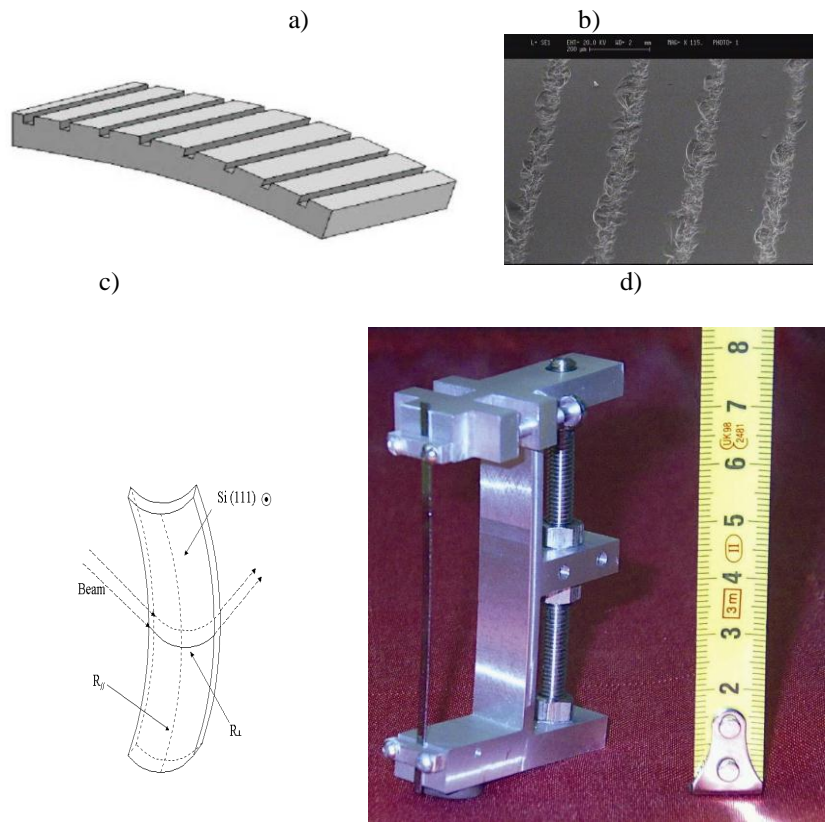

Figure 2. Methods of crystal bending.

## 3. Beam extraction from U-70

Different types of extraction schemes were realized by bent crystal. In first case high efficiency of extraction up to 85% is reached applying short silicon crystals Si 19, 22, 106 (Figure 3) . Short crystals with 2 mm in length and about 1 mrad bend take a role of first septum, than additional magnets provided deflection of circulated beam out of ring. Such high efficiency is based on multi-turn process when particles which are not captured into channeling mode on the first passage of crystal can be efficiently captured on next encounters with crystal. Extraction efficiency of 70 GeV protons versus crystal length is presented in figure 4 in comparison with simulation.

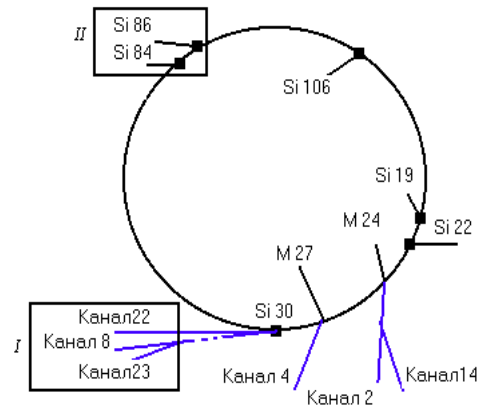

Figure 3. Crystal location at U-70 ring.

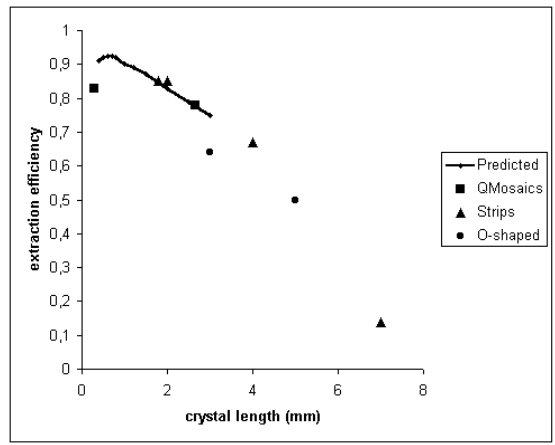

Figure 4. Efficiency dependence versus crystal length.

On Fig 5 the crystal extraction efficiency was measured as a function of proton energy. There is good agreement between the measured and calculated efficiency of bent crystal. The reduction of efficiency with reduced energy is explained by the increase of multiple scattering angle and decrease in the dechanneling length. The obtained dependence shows also that by using the same crystal one can extract beams in a broad energy range (40÷70) GeV with efficiency above 60%.

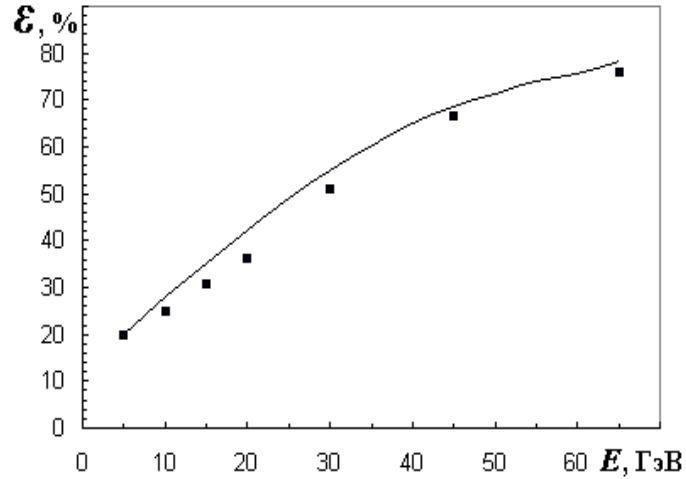

Figure 5. Crystal efficiency versus proton energy.

Important point is the crystal radiation hardness at accelerators. The limit of irradiation that channeling crystal survives was obtained in CERN and BNL experiments: ~$2\times 10^{20}$ proton per $cm^2$. Our experiments confirm these results. Crystals don't loose channeling properties during two runs 1400 hours each. As far as heat loads are concerned, our experience shows that crystal with efficiency (80÷85)% ensures beam extraction at intensity up to $10^{12}$ particle/cycle and duration (1÷2) s per cycle, suiting the requirements of the most experiments at IHEP accelerator.

With directing ~$10^{13}$ protons, crystal looses channeling properties. To define better the limit of intensity at crystal, special studies are needed including solutions for heat removal from crystal.

Proton beam extraction using bent crystals allows in principle the parallel work of several internal targets as confirmed experimentally at IHEP in 1991 [10]. Realization of this regime using short crystals opens opportunity of simultaneous work of several experimental setups on all the flattop of accelerator cycle, leading to significant reduction of expenses on the experiments. Notice that a classic resonance extraction is incompatible with the parallel work of internal targets. In this case the flattop has to be shared between them.

For creation of the regime of simultaneous work of extraction and internal targets we used crystal stations $Si_{19}$ or $Si_{106}$ (figure 3). Here, three setups could work simultaneously with the beams fully conforming to the requirements of experiments for intensity, duration and beam quality. Figure 6 shows time structure of extracted beam and of secondary beams on the flattop of cycle. The number of simultaneously performed experiments can be attain to 4 using crystal $Si_{30}$ (figure 3).

The created extraction method works at IHEP since the end of 1999 in every run of accelerator. Detailed consideration of all the systems of this extraction is given in [5].

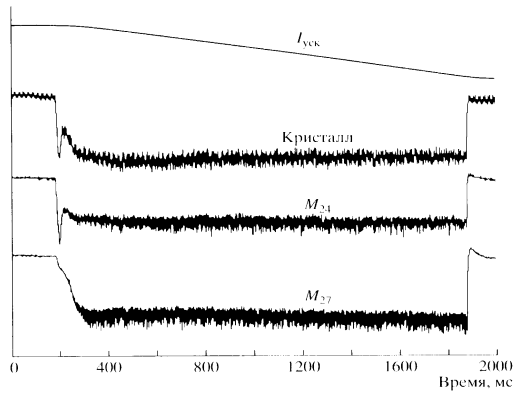

Figure 6. Monitor signals ensuring beam direction on crystals and internal targets.

Other option of extraction using long crystals (few cm in length, few tens mrad bend) was investigated. Efficiency of extraction dropped with increasing of deflection angle (Figure 7), but these moderate intensity beams are also promising for providing of physical program with protons and ions in few IHEP beamlines [11].

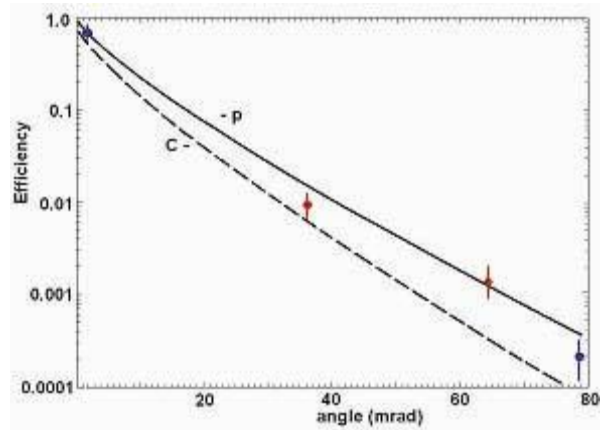

Figure 7. Efficiency dependence versus crystal bend angle.

## 4. Reflections offer new way to steer the particle trajectories

Recently IHEP group together with employees of several Russian and foreign centers of science have opened the new physical phenomenon - reflection of high energy protons from the bent atomic planes of silicon crystal ([3] and ref. herein). Volume reflection is caused by interaction of incident particle with potential of the bent atomic lattice and occurs on small length in the vicinity of a tangent to the bent atomic plane, leading to deflection of a particle aside, opposite to a bend (Figure 8). The probability of reflection is very high and comes nearer to unit at the energies bigger 100 GeV.

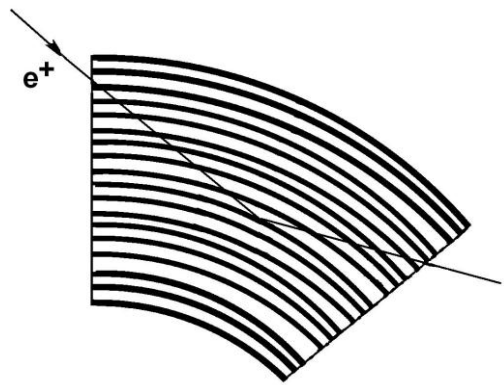

Figure 8. Scheme of volume reflection.

The phenomenon of reflection occurs in wide area of angles and is more effective, than usual channeling. Therefore there are real prospects to use of reflection for extraction and collimation of beams from the big accelerators (first of all LHC) [12]. In conditions of IHEP at U-70 creation of a highly effective extraction of beam with intensity up to $10^{13}$ in a cycle is possible due to application of multilayered reflecting structures. Calculations have shown, that 50 GeV protons can reach deflection angle sufficient for extraction from U-70 ring in case of reflections in ~10 consistently located and well oriented crystals. Such multi-crystal structures have been created in few laboratories and successfully tested in [13]. Multi-reflecting structures have been created also (figure 9) in IHEP [14].

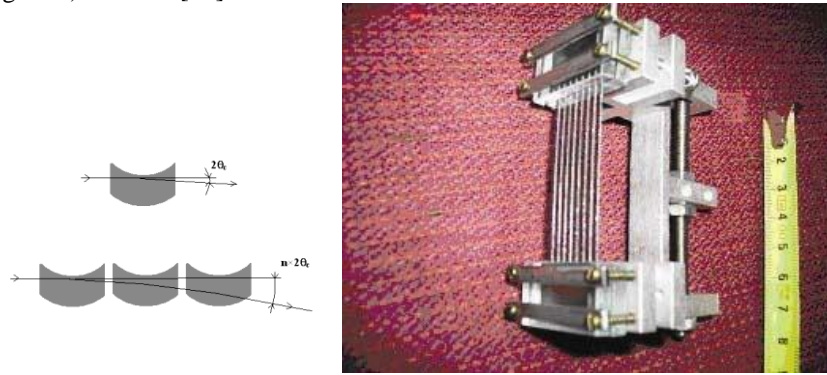

Figure 9. Enhancement of reflection angle in multicrystal.

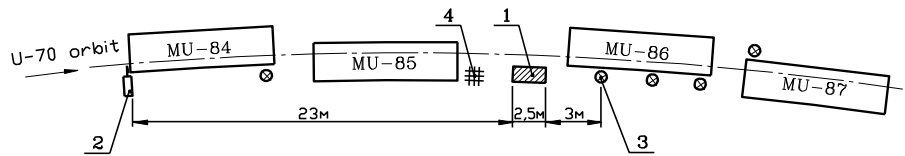

Figure 10. Scheme of crystal collimation experiment: 1-absorber, 2-crystal multireflector, 3-beam loss monitors, 4- secondary emission profilometer.

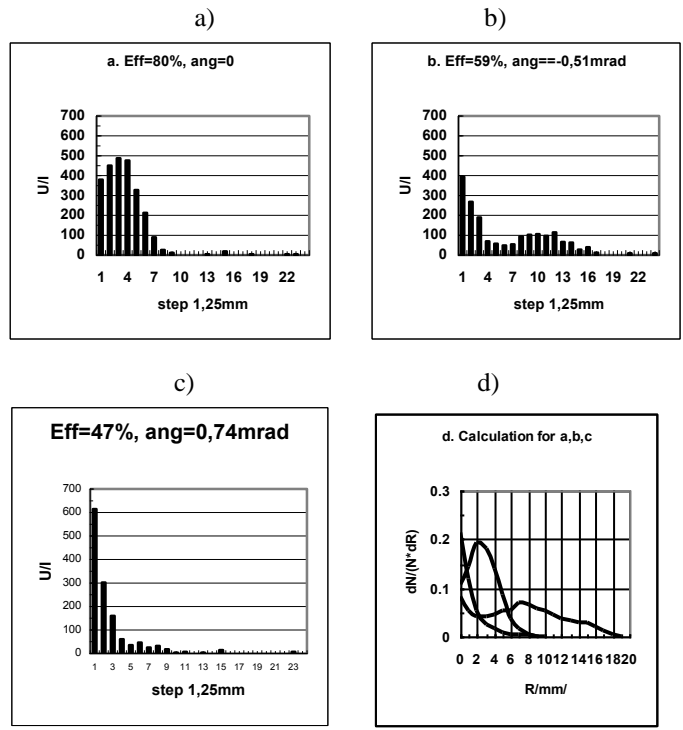

Figure 11. a), b), c) - Beam profiles at absorber edge versus crystal orientation angle. d) – expectations.

Two reflecting structures with quantity of silicon strips 10 and 7 were placed in system of localization of losses at U-70 accelerator, settling down on some distance before 3 m long steel absorber. The arrangement of crystal stations and devices of diagnostics of beam near to absorber is submitted on figure 10. On figure 11 profiles of the beam reflected by 10 strip crystal are shown. On histograms distribution of density of the deflected particles is consistently resulted at the different orientation of a crystal corresponding to: volume reflection (a), channeling (b) and disoriented crystal (c).

As seen from the figures the crystal effectively deflected circulating beam deep into the absorber body due to repeated reflection. Over 80 % of particles were reflected in a crystal and were deserted in absorber on distance over 1 mm from the edge. It is expected, that over 90 % of circulated particles can be reflected by crystal in thin-walled electrostatic deflection for extraction from the ring.

## 5. Radiation of photons from particles at volume reflection

At volume reflection particles cross a row of bent crystallographic planes (figure 8) and the behavior of its transversal velocity has nonperiodic oscillation character (figure 12). On the basis of consideration of this motion in [15] the new type of intensive coherent radiation for electrons and positrons was predicted for these conditions. Due to high value of a Lorentz factor $\gamma$ this radiation is more significant for light leptons than for other particles. Experiment on observation of this type of radiation was carried out at 22 beam line of IHEP accelerator [16]. At this beam line 10 GeV positron beam was produced. Positrons were directed on 0.65 mm silicon crystal bent at 0.5 mrad angle.

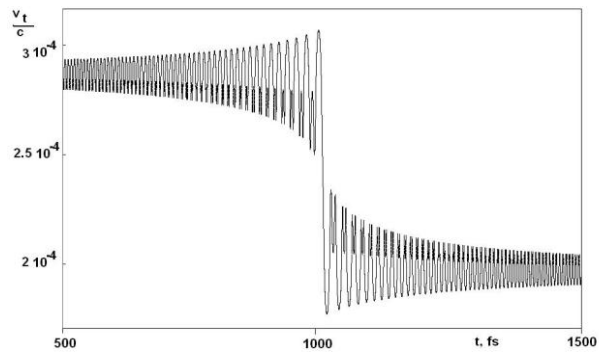

Figure 12. Behavior of velocity near reflection point.

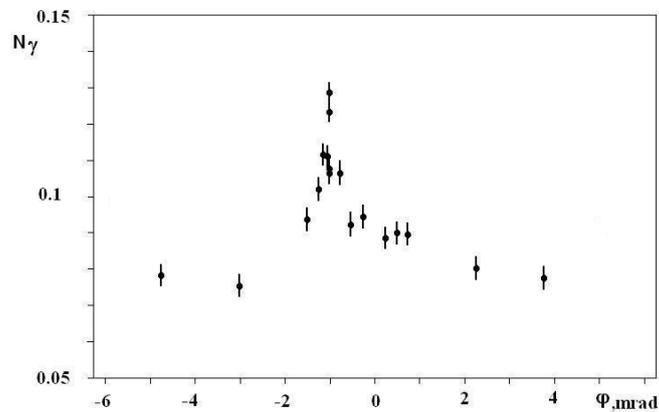

Figure 13. Orientation curve of photon yield.

11Initially the orientation dependence of radiation intensity was found. Figure 13 illustrates the result of these measurements. One can see that the orientation curve has a clear maximum. Figure 14 illustrates the measured radiation energy losses $E_\gamma dN_\gamma / dE_\gamma$ of positrons in single crystal. For comparison one can point that value of energy losses for amorphous silicon (or nonoriented crystal) is 0.007 and it is practically independent of gamma-quantum energy. Thus presented here experimental results show on valuable radiation process of 10-GeV positron beam at volume reflection in bent single crystals. Such type of radiation can be applied at accelerators for production of intensive gamma-quantum beams and positron sources. More convincing data on radiation at volume reflections were obtained recently at 180 GeV [17].

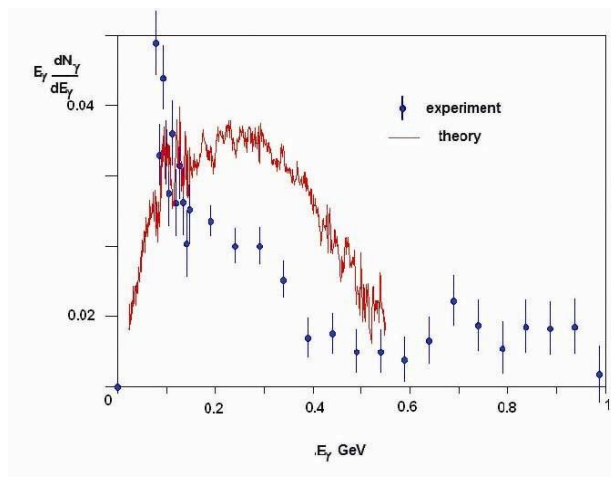

Figure 14. Differential energy losses.

This work is supported by IHEP Directorate, State corporation Rosatom and RFBR grants 07-02-00022a, 08-02-01453a, 08-02-13533ofi-c, 08-02-91020 RFBR-CERN

3. W. Scandale et al., Phys.Rev.Lett. 98 (2007) 154801

4. V.M. Biryukov, Yu.A. Chesnokov, V.I. Kotov, Crystal channeling and its application at high-energy accelerators. Berlin, Germany: *Springer* (1997) 219 pp.

5. A.G. Afonin et al., Phys.Part.Nucl. 36 (2005) 21-50

6. M.D. Bavizhev et al., *IHEP preprint* 1989-77, Protvino, (1989)

7. A.A.Arhipenko et al., Instrum.Exp.Tech. 52 (2009) 155-158

8. Afonin A. G. et al., Nucl. Instrum. Meth. B234 (2005) 122-127

9. V. M. Biryukov et al., Rev. Sci. Instrum. 73 (2002) 3170-3173

10. A.A. Aseev et al. Nucl. Instrum. Meth. A309 (1991) 1-4

11. A.A.Arhipenko et al., JETP Letters 88 (2008) 265

12. I.A. Yazynin et al. Proc. SPIE Int.Soc.Opt.Eng. 6634 (2007) 66340H

13. W Scandale et al., Phys. Rev.Lett. 102 (2009) 084801

14. A.G.Afonin et al., Atomic Energy 106 (2009) 409-414

15. Yu.A. Chesnokov et al. JINST 3 (2008) P02005

16. A.G.Afonin et al, JETP Letters 88 (2008) 486

17. W. Scandale et al., Phys.Rev. A79 (2009) 012903